# Transcritical transition of the fluid around the interface

Bonan Xu[a], Yanqi Zhu[a], Hanhui Jin[a,b]*, Yu Guo[a], Jianren Fan[b]

a. School of Aeronautics and Astronautics, Zhejiang University, Hangzhou 310027, PR China

b. State Key Laboratory of Clean Energy Utilization, Zhejiang University, Hangzhou, 310027, China

## Abstract

In this letter, we provide fundamental insights into the dynamic transcritical transition process using molecular dynamics simulations. A transcritical region, which covers three different fluid states, was discovered as a substitute for the traditional interface. The physical properties, such as temperature and density, exhibited a highly non-linear distribution in the transcritical region. Meanwhile, the surface tension was found to exist throughout the transcritical region, and the magnitude was directly proportional to $-\rho \cdot \nabla^2 \rho$.

## Main

The transcritical process occurs when subcritical fluids enter a supercritical environment, or hot-dense supercritical fluids enter a low-temperature environment. These situations widely exist in nature and industrial applications, such as eruption of submarine volcanos[1,2] and fuel injection in modern engines[3,4].

Because of the unique physical properties in the supercritical state and the significant difference in physical properties between subcritical and supercritical states, many efforts have been made to gain deeper understanding of fluids in supercritical states in

Corresponding author Hanhui Jin. Tel.: +86 057187951078; fax: +86 571 87951464; E-mail address: enejhh@emb.zju.edu.cn

the past few decades[5-7]. Researchers[8] found that the states of supercritical fluids can be differentiated into two regimes by the so-called Widom line (WL), namely the gas-like and liquid-like states. Since then, dramatic changes in the thermodynamic and transport properties between the two sides of the Widom line have been reported in several publications[9-12]. Thus, the fluid that crosses the Widom line is called transcritical fluid. Although great progress has been made in the study of transcritical states of fluids, current research mainly focuses on fluid properties at a specified state especially for the states around the Widom line, instead of the dynamic transcritical transition process of the fluid.

When the transcritical transition occurs, the properties of fluids change dramatically in the region where the local thermodynamic state crosses the Widom line, which leads to a highly nonlinear structure inside the fluid. Such non-linearities and transient evolution of the physical properties during the transcritical transition pose a great challenge to experimental technologies[10] and numerical treatments[13-15]. For example, surface tension, which was traditionally believed to disappear under supercritical pressure[4,13-17], appeared to be retained according to direct observation in more recent studies[3,18,19]. Nevertheless, no further knowledge on the surface tension in transcritical transitioning fluids is available to date.

In this study, a series of molecular dynamics (MD) simulations were conducted to investigate the transcritical transition of argon. The interaction between atoms was calculated using the Lennard-Jones 12-6 potential function ( $U = 4\epsilon[(\sigma/r)^{12} - (\sigma/r)^{12}]$ ) truncated at $3.0*\sigma$, where ε is the well depth and σ is the zero-energy separation distance. Here, we used σ = 4.5Å and $\mathcal{E}/k_B = 119.8K$ [20], and $k_B$ is the Boltzmann constant. To evaluate local surface tension, the simulation box was divided into small slabs with a width of 0.2nm in the Z direction. The surface tension was calculated as the integral of the difference between the normal and tangential components of the pressure tensor[20] .

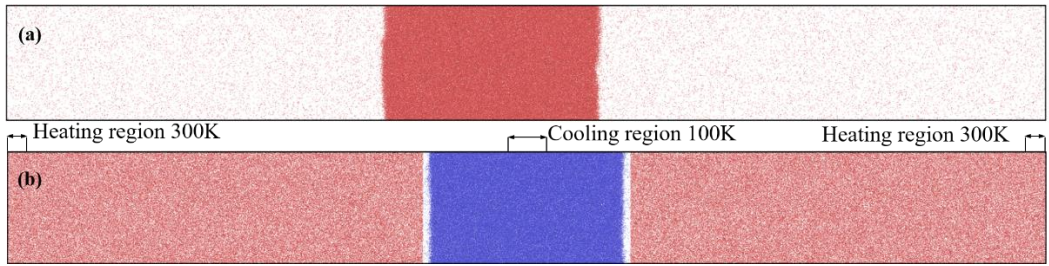

Fig. 1 Initial conditions of the MD system (a) subcritical condition (b) transcritical transition

The MD simulations were conducted in a cuboid with periodic boundary conditions in both directions, as shown in Fig.1. The co-existence of gas and liquid with an explicit interface was set for the systems under subcritical conditions at the initial stage (as shown in Fig.1a), and the simulation system was then heated to the target temperatures. For transcritical conditions, the cooling and heating zones were arranged at the middle and the boundaries of the cuboid with fixed temperatures of 100K and 300K, respectively (as shown in Fig.1b). Argon was selected as the objective fluid in the present study. Because the pressure propagation rate is much faster than that of the heat transfer, this process can be simplified as an isobaric heat transfer process[4,17,21]. In this letter, several different system pressures of P=4.905, 5.146, 5.716 and 6.257 MPa were chosen to investigate the effect of ambient conditions on the transcritical transition.

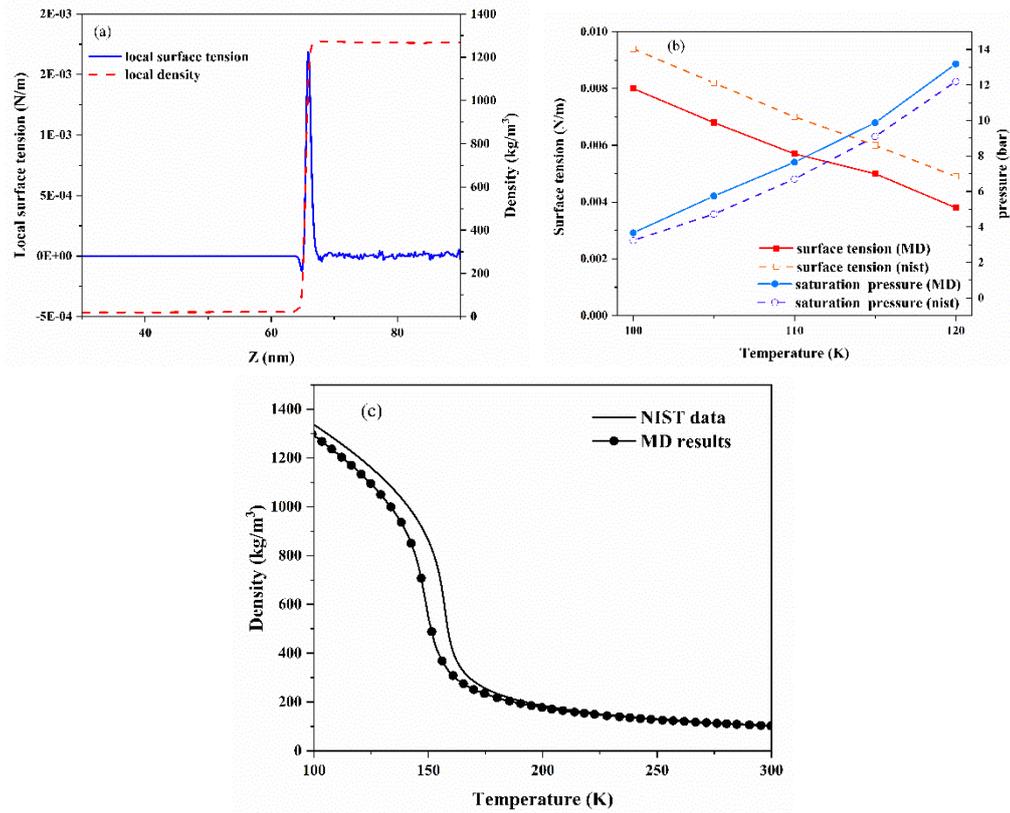

Fig. 2 Comparisons between MD results and NIST data[22] (a) Spatial distributions of density and local surface tension at subcritical condition (b) Surface tension and the saturation pressure at different temperatures (c) Comparison of density at different temperatures between NIST data[22] and MD results (P=6.2568MPa)

Fig.2 shows the MD results at subcritical states and their comparison with National Institute of Standards and Technology (NIST) data[22]. The distinct discontinuity of the density is clearly exhibited as a sudden jump at the gas-liquid interface. The surface tension, which is calculated as the integral of the difference between the normal and tangential components of the pressure tensor over the interface[20], was also obtained at the interface (as shown in Fig. 2a). Comparisons of the surface tension and saturation pressure at different temperatures were conducted between the MD results and the NIST database. The MD results were in good agreement with the standard data (Fig. 2b).

Meanwhile, the obtained densities at different temperatures under the supercritical pressure P=6.2568 MPa were compared with the NIST data[22] (as shown in Fig. 2c). The results agreed well with each other. All these comparisons indicate that the present MD simulations can properly reproduce the physical properties in both the subcritical and near-critical states.

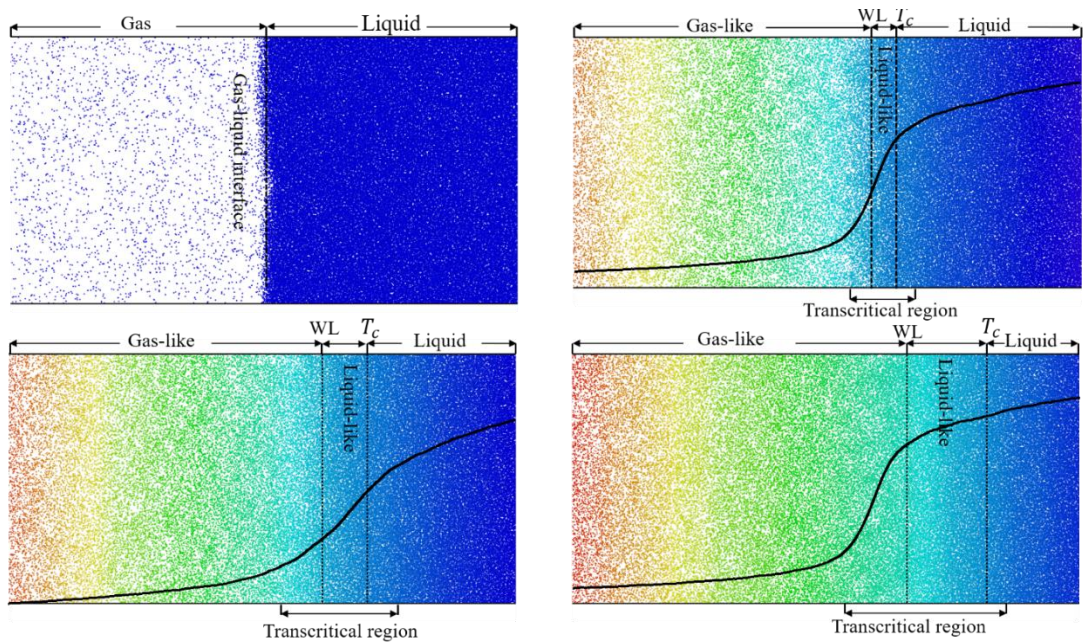

Fig. 3 Spatial distribution of the Argon molecules at different conditions

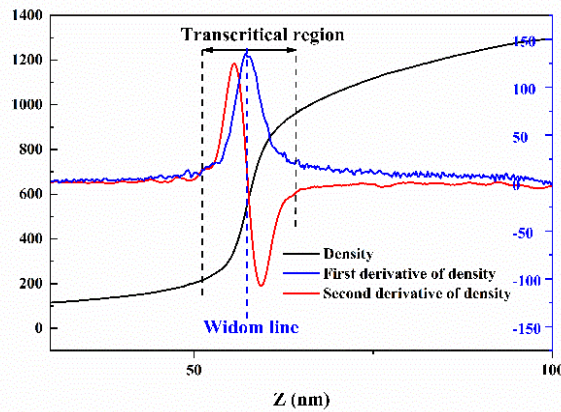

Fig.4 Spatial variation of density, first and second derivative of density inside transcritical fluid at P=5.716MPa

The spatial distributions of the Argon molecules under different conditions are shown in Fig. 3, in which the colors of the molecules represent the local temperature. The sharp saturated gas-liquid interface with density change abruptly across the interface under subcritical conditions (T=100K). The gas-liquid interface starts to dissolve in the case of transcritical conditions. Meanwhile, the blurred interface is broadened with pressure. The molecule concentration increases continuously from the supercritical gas-like region to the subcritical liquid region, but shows a large gradient around the transcritical temperature line ($T_c$). Unlike the explicitly exhibited Widom line for fluid in supercritical and subcritical states, no distinct difference in the molecule concentration can be observed in the region near the critical temperature ($T_c$) to separate the fluid into supercritical and the subcritical zones. In consideration of the commonly used definition of the maximum specific heat capacity at constant pressure as the 'Widom line' in former research to discriminate the gas-like and liquid-like states, we designate the line of largest density gradient as the 'Widom line' to spatially specify the interface between the gas-like and liquid-like zones inside the fluid (as shown in Figs. 3 and 4). It can also be observed that both the Widom line and the critical temperature line move toward the liquid side as the pressure increases. This indicates that improving the pressure can intensify the heat transfer of the gas-like liquid so that the range of the supercritical fluid rises. Meanwhile, the widths of the gas-like and liquid-like zones are extended with the pressure improvement simultaneously. This implies that the passivation and broadening of the interface are concurrent with the transcritical transition inside the fluid. The higher the pressure, the more intense the passivation and broadening. This phenomenon can also explain the quick completion of a single droplet's transcritical transition at very high temperatures and pressures. At the Widom line, the gradient of density reaches its maximum value, namely the first derivative of the local density $d\rho/dx$ reaches its maximum and $d^2\rho/dx^2 = 0$ (as shown in Fig. 4). The region of $d^2\rho/dz^2 \neq 0$ is defined as the transcritical region.

Fig.5 shows the specific spatial variations of temperature (T) density (ρ), derivative of density ($d\rho/dz$), local surface tension (γ), and $-\rho \cdot d^2\rho/dz^2$ at different pressures. The nonlinear distributions of temperature and density can be clearly observed. Owing to the influence of heat transfer and nonlinear distributed temperature (as shown in Fig.5a), the density of the fluid increases monotonically in the Z direction as the state changes from a gas-like state to a liquid state (as shown in Fig. 5b). The Widom line is defined as the location of the largest density gradient (as shown in Fig. 5c). Meanwhile, other physical properties, such as the surface tension, also show highly non-linear distributions around the Widom line (as shown in Fig. 5d). The appearance of local surface tension (LST) becomes zero at the Widom line, which implies consistency between the Widom line for spatial distribution in present paper and the traditional Widom line for the fluid state. Thus, the first derivative of the fluid density $d\rho/dx$ can be used to define the location of the Widom line. The surface tension can be found to exist throughout the transcritical region instead of disappearing in the supercritical environment in previous assumptions[17,21]. Compared to that at subcritical conditions, the local surface tension exists in a much wider zone with a lower magnitude. This is because of the non-linear distribution of the molecule concentration in the transcritical region, which leads to non-equilibrium interactions between molecules. The span of the local surface tension increases with the system pressure, which is consistent with the broadening of the transcritical region when the pressure is improved. Furthermore, it can also be observed that the magnitude of surface tension varies in sync with the variable $-\rho \cdot d^2\rho/dz^2$ (as shown in Fig.5e). This implies that the surface tension mainly depends on the local density and its second derivative. As a result, the nonlinear distribution zone of the density for which $d^2\rho/dz^2 \neq 0$ can be utilized to define the transcritical region in the fluid. The local surface tension exhibits different variation tendencies between different sub-regions because of the nonlinear distribution of the density. The negative surface tension, which was found in the gas next to the gas-liquid interface under subcritical conditions[23], is observed in the gas-like zone. At the Widom line, the local surface tension returns to zero and then starts to increase in the liquid-like zone. In the liquid-like and liquid zones, it appears positive and then gradually returns to zero at the position of $d^2\rho/dz^2 = 0$, namely the other end of the transcritical region.

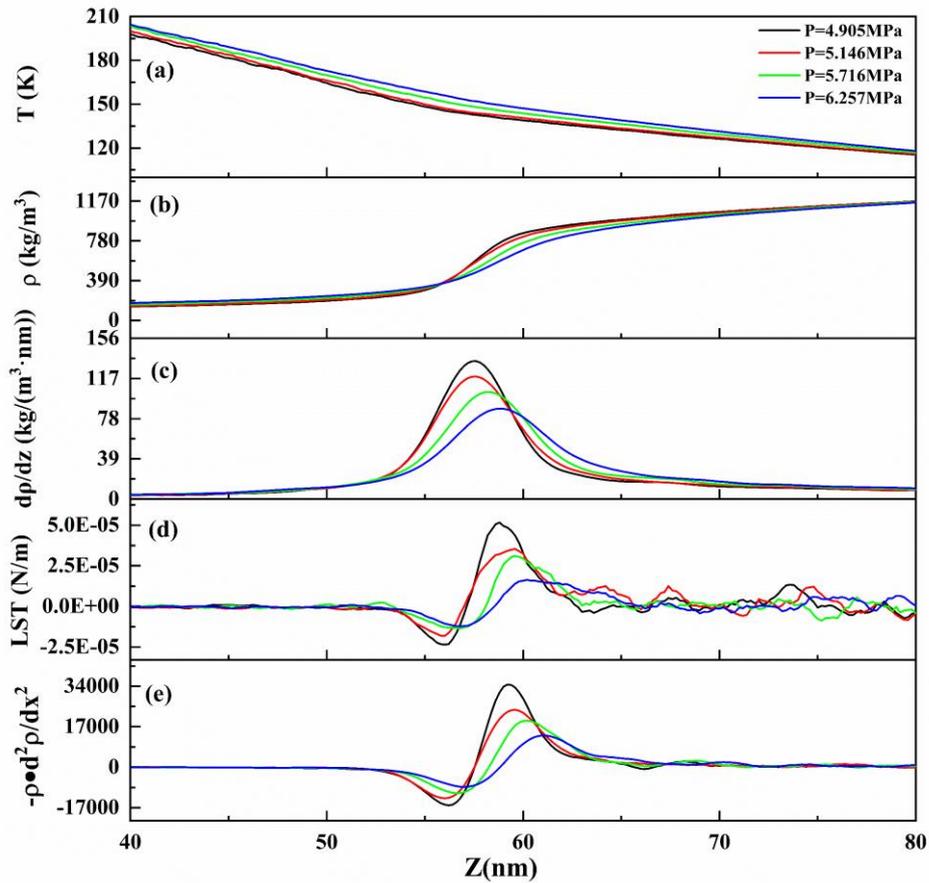

Fig. 5 Spatial variation of density, derivative of density, surface tension and $\rho \cdot d^2\rho/dz^2$ (a) temperature (b) density (c) derivative of density (d) local surface tension (e) $-\rho \cdot d^2\rho/dz^2$

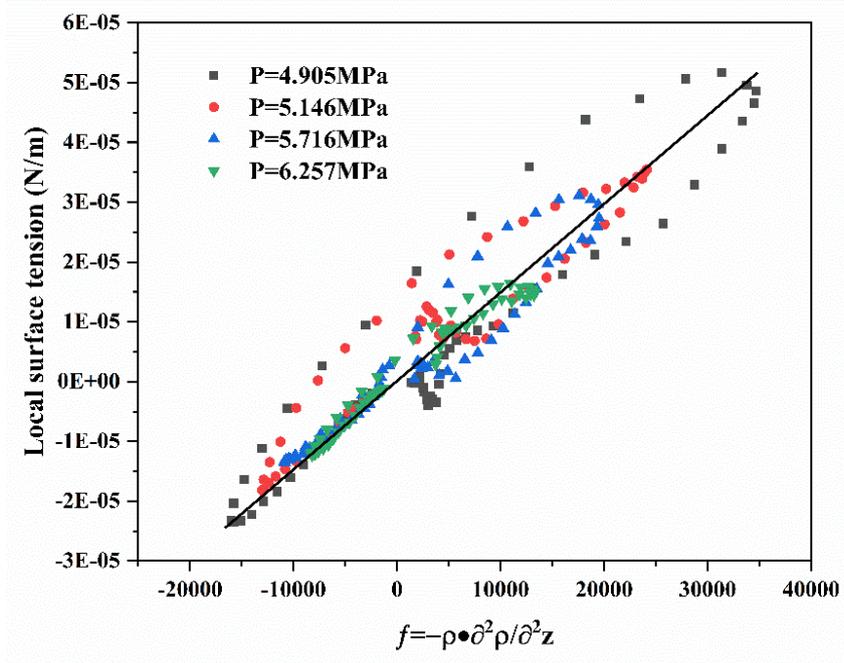

Fig. 6: The relationship between surface tension and the variable $-\rho \cdot d^2\rho/dz^2$

Further investigation shows that the local surface tension throughout the transcritical region is almost directly proportional to the variable $-\rho \cdot d^2\rho/dz^2$ (as shown in Fig.6)

$$\gamma = -c\rho \cdot \nabla^2 \rho \qquad (1)$$

where, γ is the local surface tension, c is a constant (c=1.5e-8 is obtained for argon in the present research), ρ is the local density of the fluid, which represents the effect of local concentration of the molecules, and $\nabla^2$ is the Laplacian operator and $\nabla^2 \rho$ represents the effect of the nonlinear distribution of the molecules across the transcritical region.

In this letter, the transcritical transition of fluid was numerically investigated. The transcritical region, which can be divided into sub-regions of gas-like, liquid-like and liquid, was discovered and defined as the region where $d^2\rho/dz^2 \neq 0$. The Widom line, which discriminates between liquid-like and gas-like zones inside the fluid, was redefined as the position where the maximum density gradient occurs. The nonlinear distribution of the physical properties such as the temperature and density were disclosed in the transcritical region. The surface tension was found to exist throughout the transcritical region instead of disappearing in supercritical environment in former assumption. Meanwhile, the local surface tension was found to be in directly proportional to $-\rho \cdot \nabla^2 \rho$. These findings provide new insights and basic knowledge that can help further understand the physics of fluids experiencing dynamic transcritical transition process.

## Acknowledgements

The authors gratefully acknowledge the financial support of the Natural Science Fund of China (Grant No: 91741103) and the National Key Research and Development Program of China (Grant No: 2016YFB0600101).

# Reference:

[1] P. F. McMillan and H. E. Stanley, Nature Physics **6**, 479 (2010).
[2] D. S. Kelley, J. A. Baross, and J. R. Delaney, Annual Review of Earth and Planetary Sciences **30**, 385 (2002).
[3] L. Jofre and J. Urzay, Progress in Energy and Combustion Science **82**, 100877 (2021).
[4] V. Yang, Proceedings of the Combustion Institute **28**, 925 (2000).
[5] R. Shi and H. Tanaka, Proceedings of the National Academy of Sciences **117**, 26591 (2020).
[6] B. Cheng, G. Mazzola, C. J. Pickard, and M. Ceriotti, Nature **585**, 217 (2020).
[7] V. P. Sokhan, A. Jones, F. S. Cipcigan, J. Crain, and G. J. Martyna, Physical Review Letters **115**, 117801 (2015).
[8] G. G. Simeoni, T. Bryk, F. A. Gorelli, M. Krisch, G. Ruocco, M. Santoro, and T. Scopigno, Nature Physics **6**, 503 (2010).
[9] P. Sun, J. B. Hastings, D. Ishikawa, A. Q. R. Baron, and G. Monaco, Physical Review Letters **125**, 256001 (2020).
[10] F. Maxim, C. Contescu, P. Boillat, B. Niceno, K. Karalis, A. Testino, and C. Ludwig, Nature Communications **10**, 4114 (2019).
[11] P. Gallo, D. Corradini, and M. Rovere, Nature Communications **5**, 5806 (2014).
[12] F. Gorelli, M. Santoro, T. Scopigno, M. Krisch, and G. Ruocco, Physical Review Letters **97**, 245702 (2006).
[13] C. Rodriguez, A. Vidal, P. Koukouvinis, M. Gavaises, and M. A. McHugh, Journal of Computational Physics **374**, 444 (2018).
[14] P. C. Ma, Y. Lv, and M. Ihme, Journal of Computational Physics **340**, 330 (2017).
[15] S. Kawai, H. Terashima, and H. Negishi, Journal of Computational Physics **300**, 116 (2015).
[16] P. E. Lapenna and F. Creta, The Journal of Supercritical Fluids **128**, 263 (2017).
[17] P. E. Lapenna, Physics of Fluids **30**, 077106 (2018).
[18] C. Crua, J. Manin, and L. M. Pickett, Fuel **208**, 535 (2017).
[19] J. Manin, M. Bardi, L. M. Pickett, R. N. Dahms, and J. C. Oefelein, Fuel **134**, 531 (2014).
[20] M. J. P. Nijmeijer, A. F. Bakker, C. Bruin, and J. H. Sikkenk, Journal of Chemical Physics **89**, 3789 (1988).
[21] H. Meng, G. C. Hsiao, V. Yang, and J. S. Shuen, Journal of Fluid Mechanics **527**, 115 (2005).
[22] NIST Chemistry WebBook, https://webbook.nist.gov/chemistry/.
[23] S. H. Park, J. G. Weng, and C. L. Tien, International Journal of Heat and Mass Transfer **44**, 1849 (2001).